\documentclass[pre,showpacs,twocolumn,letter]{revtex4-1}
\usepackage{color}
\usepackage{epsfig}
\usepackage{amssymb,amsmath,amsthm}
\usepackage{esint}
\usepackage[colorlinks=true,linkcolor=blue,citecolor=blue]{hyperref}
\usepackage{graphicx}
\definecolor{red}{rgb}{1,0,0}

\begin{document}
\title{Quasi-reversible parametric instability in presence of noise}
\author{Marcel G. Clerc}
\email{marcel@dfi.uchile.cl}
\author{Claudio Falc\'on}
\email{cfalcon@uchile.cl}
\affiliation{Departamento de F\'isica, Facultad de Ciencias F\'isicas y Matem\'aticas, Universidad de Chile, Casilla 487-3, Santiago, Chile}
\author{Ren\'e G. Rojas}
\affiliation{Instituto de F\'isica, Pontificia Universidad Cat\'olica de Valpara\'iso,
Casilla 4059, Valpara\'iso, Chile}

\date{\today}

\begin{abstract}

We present an experimental and theoretical study of the effect of spatio-temporal fluctuations in quasi-reversible systems displaying a spatial quintic supercritical bifurcation. The saturation mechanism is drastically changed by the inclusion of fluctuations. Experimentally, we observe the modification of the bifurcation diagram of parametrically amplified surface waves as spatiotemporal fluctuations stemming from an underlying vortex flow are included. Theoretically, we characterize the noise-dependent effective dynamics in a model system, the parametrically driven nonlinear Schr\"odinger equation, subjected to noise which allows us to rationalize the effect of the underlying vortex flow on the surface waves.

\end{abstract}


\maketitle

In nature, most of physical systems present fluctuations. These fluctuations are commonly discarded or considered as a nuisance of the signal under study. Even so, fluctuations can have rather surprisingly constructive and counterintuitive effects. The most well-known examples in zero dimensional systems are noise induced transitions~\cite{Lefever} and stochastic resonance~\cite{Hangii}. More recently, examples of the effect of noise in spatially extended systems have been studied, such as noise-induced phase transitions and patterns~\cite{OS99}, noise-sustained structures in convective instabilities~\cite{AGTL06}, and front propagation~\cite{ClercNoise2006}, to mention a few. A counterintuitive consequence of the effect of noise is the modification of the deterministic bifurcation curves of order parameters, changing their critical exponents and bifurcation points. Studying the manner in which the deterministic growth of an order parameter depends on the statistical properties of noise can give insight into the nonlinear amplification mechanism or the internal dynamics of the system under study~\cite{FDT}. From this point of view it is relevant to characterize the role of fluctuations in the deterministic bifurcation curves~\cite{NoiseSupercritical}, specially when injection or dissipation of energy are almost negligible as in the case quasi-reversible systems~\cite{QuasiReversal}. Generically, close to the parametric instability onset, this type of systems displays a quintic supercritical transition which corresponds to a bifurcation between super and subcritical instabilities, where noise drives the bifurcation. This type of saturation has already been observed in Ref.~\cite{FauvePetrelis2005}.

Here we examine the effect of fluctuations on a supercritical bifurcation with a quintic nonlinearity as the saturation mechanism. Experimentally, standing waves at the surface of mercury are parametrically amplified by the vertical modulation of gravity and subjected to spatio-temporal fluctuations by means of an underlying vortex flow. From a prototype model used to describe the surface displacement and the velocity potential of the fluid---parametrically driven damped nonlinear Schr\"odinger equation~\cite{PDNLS}---we deduce the amplitude equation for the standing waves~\cite{PDNLSFaraday}, which satisfies a supercritical real Ginzbug-Landau equation with quintic nonlinearity and multiplicative noise. Controlling fluctuations by means of an underlying vortex flow alters drastically the threshold, the type and the shape of the bifurcation. We charaterize this changes and contrast them with theory.

\begin{figure*}[t]
\includegraphics[width=1.0\columnwidth]{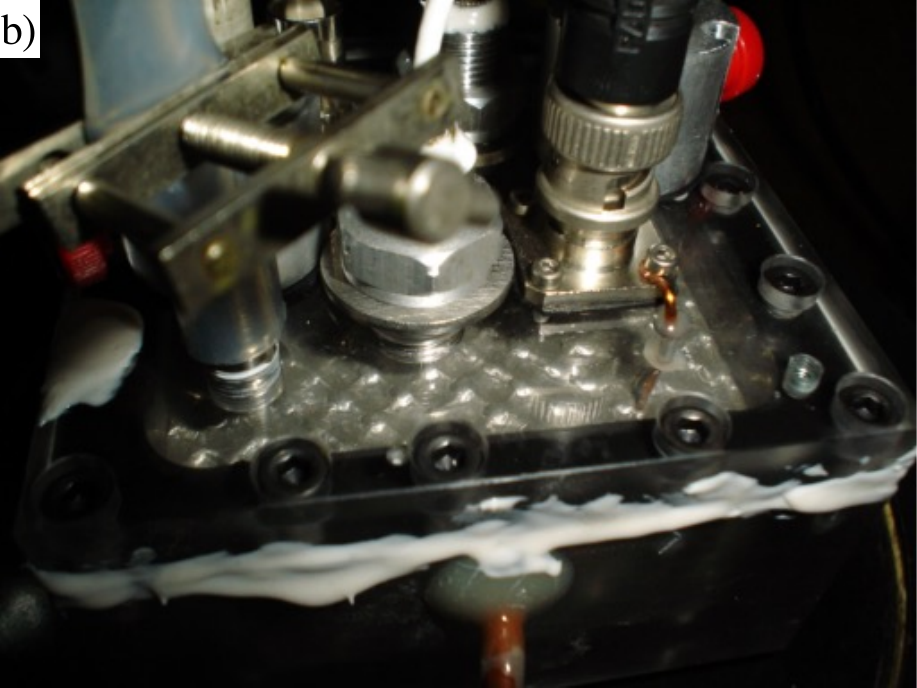}
\includegraphics[width=1.0\columnwidth]{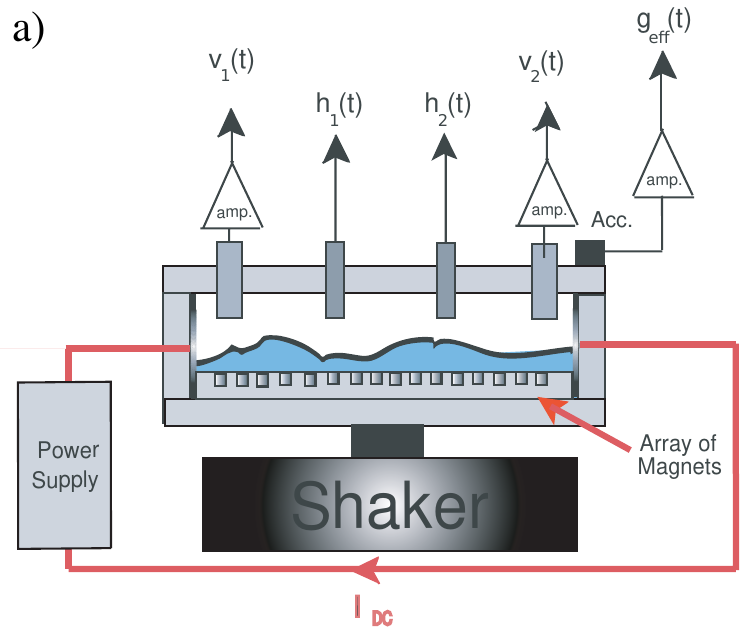}
\caption{{\bf Left}: View of the $10\times10\times5$ cm$^{3}$ experimental setup. {\bf Right}: Schematic diagram of the experimental set-up. A $5$ mm thin mercury layer is enclosed in a Plexiglass box and vibrated by an electromechanical shaker. Surface amplitude is measured by inductive sensors. }
\label{F-fig1}
\end{figure*}
The experimental set-up is already described in Ref.~\cite{Falcon2009} and shown in Fig.~\ref{F-fig1}. A plexiglass container (7$\times$7 cm$^{2}$) filled with a 5 mm layer of mercury is mounted over an electromagnetic vibration exciter (B$\&$K 4809), driven by a frequency synthesizer (HP 8904A), providing a clean vertical sinusoidal acceleration (horizontal acceleration less than 1 $\%$ of the vertical one). The sinudoidal gravity modulation $g_{eff}(t)=acos(2\pi f_{ex} t)$ is measured by a piezoelectric accelerometer (B$\&$K 4383) and a charge amplifier (B$\&$K 2635), where $a$ is proportional to the applied tension with a 1.0 Vs$^{2}$/m sensitivity and $f_{ex}$ is the excitation frequency. The surface is kept clean in a nitrogen-filled atmosphere and is temperature-regulated by circulating water 20.0 $\pm$ 0.1$^\circ$C. At the bottom of the cell, alternating vertical polarity magnets (diameter $\phi$=5 mm) are placed in an hexagonal array (wavelength $\lambda$=6 mm). The magnetic field strength at the surface of the fluid on top of a magnet is $500$ G. Two nickel-barnished copper electrodes are glued at opposite sides of the cell. Through these electrodes, a DC current $I$ generated by a power supply (Agilent E3634A) is applied giving rise to a horizontal current density ${\bf j}$ of the order of 10$^{2}$ Am$^{-2}$ and a periodic Lorentz force ${\bf F}_L={\bf j}\times{\bf B}$. Measurements of local wave amplitude are performed by two inductive sensors (eddy-current linear displacement gauge, Electro 4953 sensors with EMD1053 DC power supply, 3 mm in diameter). Additional measurements of the fluid local velocity field are performed close to the walls by two home made Viv\`es probes~\cite{Vives}. The linear sensing range of both sensors has already been checked in Ref.~\cite{Falcon2009}. The local height, acceleration and velocity fluctuations are sampled at 500 Hz with an acquisition time $T=$ 800 s.

\begin{figure}[b]
\includegraphics[width=1.0\columnwidth]{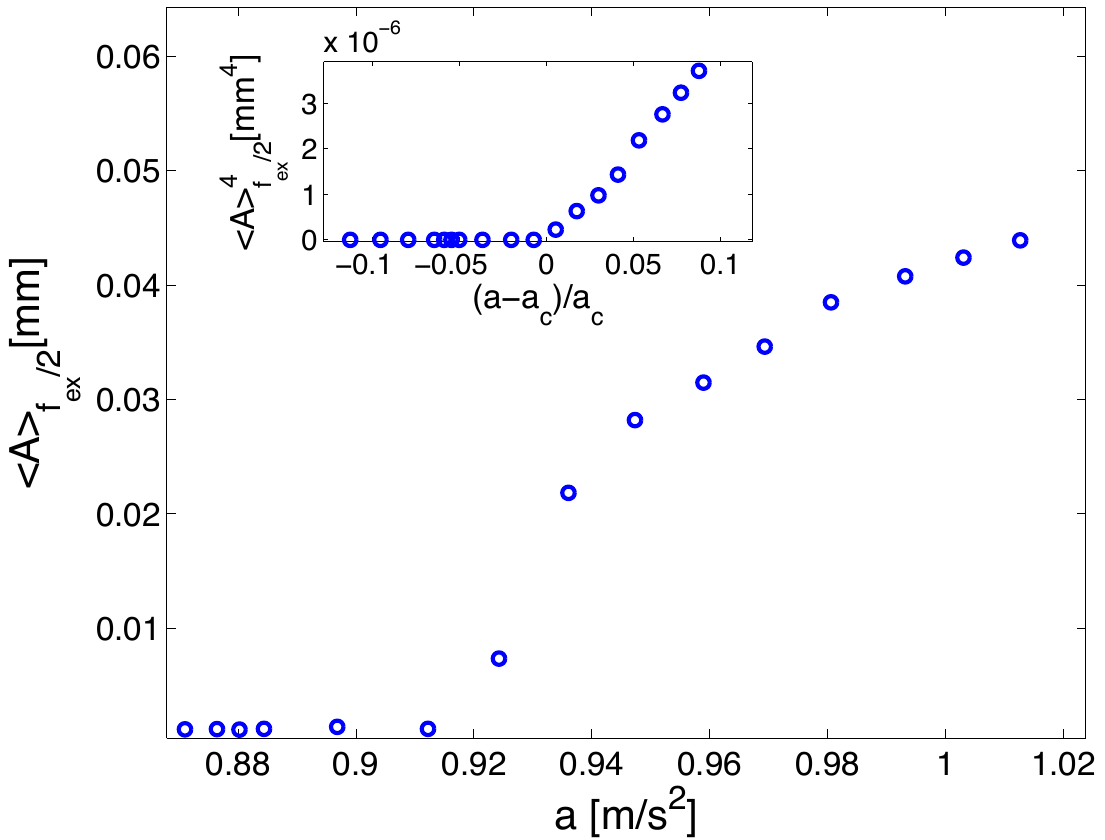}
\caption{Bifurcation diagram of $\langle A \rangle_{f_{ex}/2}$ as a function of the gravity modulation amplitude $a$. Inset: Bifurcation diagram of the fourth power of $\langle A \rangle_{f_{ex}/2}$ as a function of the normalised control parameter $(a-a_c)/a_c$.}
\label{F-fig2}
\end{figure}

For a vibration frequency $f_{ex}$ of 23.8 Hz, there is a critical acceleration value $a_c=$0.93 ms$^{-2}$ above which standing surface waves develop over the flat interface from small perturbations. These parametrically amplified waves grow over the whole container oscillating at $f_{ex}/2$ in a square pattern with wavelength $\Lambda=$ 7 mm $\sim\lambda$. No defects are observed over the wave pattern~\cite{Douady1990}. From the recorded signals we compute the Fourier amplitude of the field at $f_{ex}/2$

\begin{equation}
\left<A\right>_{f_{ex}/2}=\lim_{T\rightarrow\infty}\left|\frac{1}{2T}\int_{-T}^{T}A(t)e^{\pi i f_{ex} t}dt\right|.
\label{EqFourier}
\end{equation}
The dependence of $\langle A \rangle_{f_{ex}/2}$ on $a$ is displayed in Fig.~\ref{F-fig2}. Close to onset, for $a>a_c$, the unstable mode saturates as $(a-a_c)^{\alpha}$ with $\alpha=1/4$. This is confirmed in the inset of Fig.~\ref{F-fig2}, where $\langle A \rangle^{4}_{f_{ex}/2}\propto\epsilon$ for $\epsilon=(a-a_c)/a_c>0$ and zero otherwise. No distinguishible hysteresis is observed. From bifurcation theory, a supercritical bifurcation generically satisfies a law $\langle A \rangle_{f_{ex}/2}\propto\epsilon^{1/2}$, which is a consequence of cubic nonlinearities for the pattern amplitude equation~\cite{CrossHohenberg}. Indeed, the supercritical bifurcation diagram shown in Fig.~\ref{F-fig2} is due to the saturation of surface waves by quintic nonlinearities. Hence, the Faraday instability in this parameter range exhibits a quintic supercritical bifurcation, which is a supercritical bifurcation leaded by quintic nonlinearities.

\begin{figure}[b]
\includegraphics[width=1.0\columnwidth]{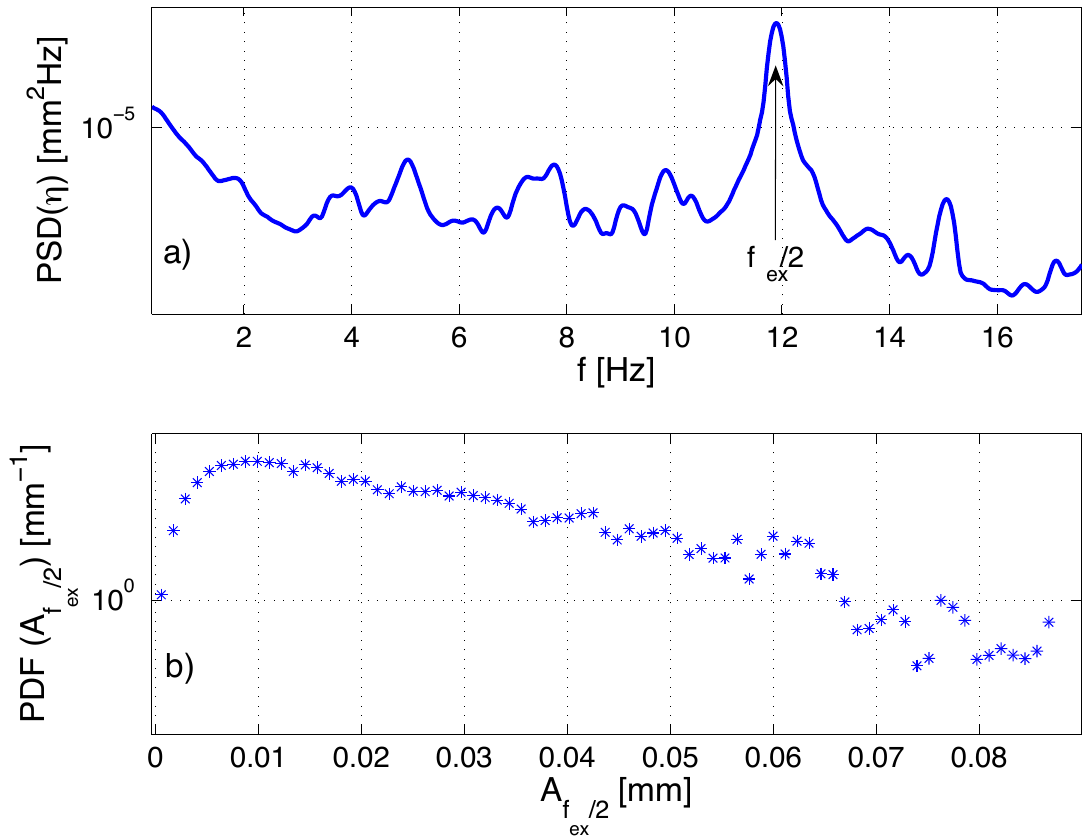}
\caption{a) Power spectral density (PSD) of the local heigh fluctuations $\eta(t)$ and b) probability density function (PDF) of the subharmonic waves envelope $|A_{f_{ex}/2}|$ for $a=$ 0.115 ms$^{-2}$ and $I=$ 1.0 A.} 
\label{F-fig3}
\end{figure}
Interaction with a vortex flow, induced by ${\bf F}_L$, creates spatio-temporal fluctuations over the standing wave pattern. Local height fluctuations $\eta(t)$ have been characterized by their power spectral density (PSD) and probability density function (PDF) in Ref. \cite{Falcon2009} in the case of purely electromagnetic forcing. When parametric excitation is added, the amplitude fluctuations of the subharmonic waves $A_{f_{ex}/2}$ is a fluctuating quantity also. We show in Fig.~\ref{F-fig3}-a) the PSD of the local height fluctuations for $a=$ 0.115 ms$^{-2}$ and $I=$ 1.0 A in the case described above. Low-frequency fluctuations dominate for $f<$3 Hz and there is a band of excited frequencies $\Delta f\sim $ 1 Hz close to the parametric resonance at $f=f_{ex}/2$. Taking this frequency band into account we calculate the PDF of the envelope $|A_{f_{ex}/2}|$, which is still fluctuating, as shown in Fig.\ref{F-fig3}-b). An exponential tail is observed for values of $|A_{f_{ex}/2}|$ larger than its mean. A steep descent for small values of $|A_{f_{ex}/2}|$ shows that almost no defects are created by the vortex flow.

\begin{figure}[b]
\includegraphics[width=1.0\columnwidth]{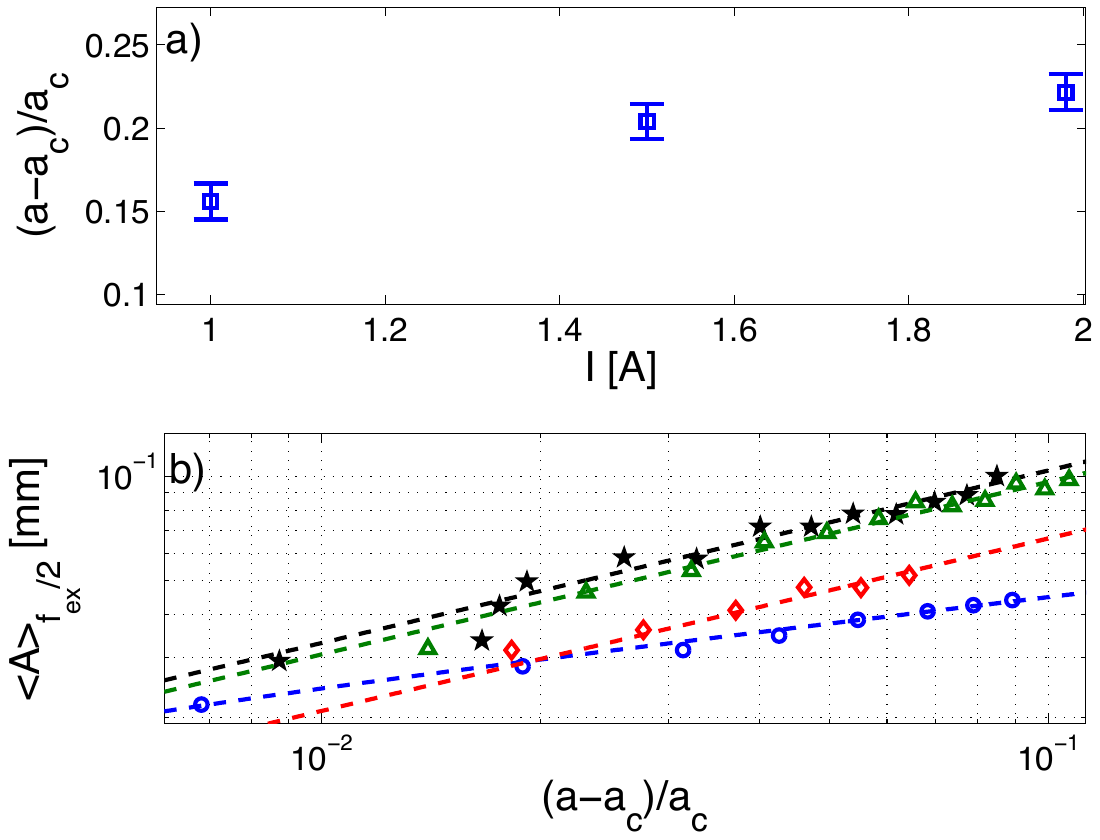}
\caption{a) Normalized bifurcation point $(a_c(I)-a_c)/a_c$ of $\langle A \rangle_{f_{ex}/2}$ as a function of $I$.  b) Loglog plot of the bifurcation curves of $\langle A \rangle_{f_{ex}/2}$ for $I=$ 0.0 ($\circ$), 1.0 ($\diamond$), 1.5 ($\star$) and 2.0 ($\triangle$) A. Dashed lines represent best fit curves.} 
\label{F-fig4}
\end{figure}

In our experimental set-up, the nonlinear interaction of vortex flow and parametric waves creates several effects over the deterministic bifurcation curves of $\langle A \rangle_{f_{ex}/2}$. There are two main effects of fluctuations in this system. First, the curent-dependent critical acceleration $a_c(I)$ or normalized bifurcation points $(a_c(I)-a_c)/a_c$ are shifted to higher values with incresing current $I$, as shown in Fig.~\ref{F-fig4}-a). Second, the saturation exponents steepen from 1/4 for purely parametric forcing to 1/2 when the vortex flow is active, as shown in the inset of Fig.~\ref{F-fig4}-b). These two effects are a clear signature of the nonlinear interaction of the vortex flow with the parametrically amplified surface waves. 

To explain the previous bifurcation diagram change, and how noise affects it, we consider that this instability is a quasi-reversal one. Thus, it is well modeled by a system that is invariant under time reversal but which is perturbed by energy injection and dissipation~\cite{QuasiReversal}. Then, a layer of thickness $h$ of incompressible fluid that is driven by a sinusoidal force normal to the free surface $z=0$ is modeled by the dimensionless parametrically forced and damped nonlinear Schr\"{o}dinger equation~\cite{PDNLSWater},
\begin{eqnarray}
\partial _{t}\psi &=& (-i\nu-\mu) \psi -\gamma \bar{\psi} \nonumber \\ &-& i|\psi|^{2}\psi -i\partial _{xx}\psi +\sqrt{\Delta}\xi(\psi,t,x),
\label{E-PDNLS}
\end{eqnarray}
where $\psi(x,t)$ is a complex field, and $\bar{\psi}$ stands for complex conjugate of $\psi$. The surface displacement from the flat surface $\eta(x,t)$ and the velocity potential at the free surface $\phi(x,z,t)\vert_{z=\eta}$ are slave variables of the form $h(x,t)=\psi(x,t) e^{-i\pi f_{ex} t}+c.c.$ and $\phi(x,z,t)\vert_{z=\eta}=-i\psi(x,t) e^{-i\pi f_{ex} t}+c.c.$, respectively~\cite{PDNLS}. $\nu$ is the detuning parameter, which is proportional to the difference between the observed standing wave frequency and $\pi f_{ex}$. $\mu$ is the damping coefficient which is proportional to the fluid kinematic viscosity in the small viscosity limit~\cite{Ursell} and $\gamma$ is proportional to $a$. $\xi$ is a stochastic function that models the large number of eliminated fast variables with intensity $\Delta$. For $\mu=\gamma=\Delta=0$, Eq.~(\ref{E-PDNLS}) becomes the one-dimensional nonlinear Schr\"odinger equation~\cite{Newell}, which describes the envelope of an oscillatory system. Expansion of Eq.~(\ref{E-PDNLS}) to two or three dimensional systems is straightforward. This model is a time-reversal Hamiltonian system with the transformation $\{ t \rightarrow -t,\psi \rightarrow \bar{\psi} \}$. The terms proportional to $\mu$ and $\gamma$ break the time reversal symmetry, and represent energy dissipation and injection, respectively. Higher order terms in Eq. (\ref{E-PDNLS}) are ruled out by scaling analysis, since $\mu \ll 1$, $\nu\sim \mu \sim \gamma$, $|\psi| \sim \mu^{1/2}$, $\partial_x \sim \mu^{1/2}$, and $\partial_t \sim \mu^{1/2}$.

Model (\ref{E-PDNLS}) has also been used to describe patterns and solitons in several other systems, such as localized structures in nonlinear lattices~\cite{NonLinearLatices}, light pulses in optical fibers~\cite{OpticalFiber}, Kerr-type optical parametric oscillators~\cite{OPO}, magnetization solitons in easy-plane ferromagnetic exposed to an oscillatory magnetic field~\cite{Magnetization,ClercCoulibalyLaroze2008}, and a parametrically driven chain of pendula~\cite{ChainPendula}, to mention a few.

A trivial state of equation (\ref{E-PDNLS}) is the homogeneous state $\psi_0=0$, which represents the flat and quiescent solution of the  fluid layer. For negative detuning, $\nu<0$, the $\psi_0=0$ state becomes unstable through a subcritical stationary instability at $\gamma^2=\mu^2+\nu^2$, which corresponds to a subharmonic instability of the flat fluid layer. Inside this region---Arnold tongue---the system has three unstable uniform solutions $\psi_0=0$, and $\psi_\pm=x_0\pm i \sqrt{(\mu-\gamma)/(\mu+\gamma)} x_0$, where $x_0\equiv\sqrt{(\gamma-\mu)(-\nu+\sqrt{\gamma^2-\nu^2})/2\gamma}$. These three states merge together through a pitchfork bifurcation at $\gamma^2=\mu^2+\nu^2$, with $\nu>0$. However, for possitive detuning the quiescent state is only stable for $\gamma<\mu$, because this state exhibits a spatial instability at $\gamma=\mu$, which gives rise to a spatial periodic state with wave number $k_c=\sqrt{\nu}$. This state represents stable subharmonic surface waves. In order to study this spatial instability, we introduce the change of variable for positive detuning 
\begin{eqnarray}
\left(
\begin{array}{c}
\psi \\
\bar{\psi}
\end{array}
\right) &=& A e^{ik_cx} \left(
\begin{array}{c}
1/2\\
0
\end{array} \right)-\frac{3}{4\mu}|A|^2Ae^{ik_cx} \left(
\begin{array}{c}
0\\
1/2
\end{array} \right)+ \nonumber \\
&&+\frac{A^3}{32\nu}e^{ik_cx}\left(
\begin{array}{c}
1/2\\
0
\end{array} \right)
+c.c+h.o.t.
\label{E-ChangeVariable}
\end{eqnarray}
The field $A(x,t)$ is the amplitude of Faraday waves, and it satisfies the quintic supercritical amplitude equation \cite{CoulletFrischSonnino}
\begin{equation}
\partial _{t}A=\Gamma A-\frac{9}{16\mu}|A|^{4}A+\partial _{xx}A
+ A\zeta(A,x,t), \label{E-DegenerateSupercritical}
\end{equation}
where $\Gamma\equiv \gamma-\mu\propto\epsilon\ll 1$ is the bifurcation parameter of the spatial instability and $\zeta$ is a stochastic term obtained from the stochastic Eq.~(\ref{E-PDNLS}) using the standard normal form method (see appendix A in Ref.~\cite{ClercFalconTirapegui}). By generic arguments of stochastic normal forms this term can be modeled as the sum of gaussian white noise functions, i.e. $\zeta(A,x,t)=\sum_{i=1,j=1} \sqrt{\Delta_{i,j}}\zeta_{i,j}(x,t)A^{i}\bar{A}^{j}$, with zero mean value and correlation $\left\langle \zeta_{i,j} \left( x,t\right) \zeta_{p,q} \left( x^{\prime },t^{\prime }\right) \right\rangle =\delta \left( x-x^{\prime }\right) \delta \left( t-t^{\prime }\right) \delta_i^p \delta_j^m$. Here the terms  $\Delta_{i,j}\propto\Delta $ represent the noise intensity for the different multiplicative white noise terms $\zeta_{i,j}(x,t)$.

The deterministic model (\ref{E-DegenerateSupercritical}) has a family of periodic solutions of the form $A(x,t)=2\sqrt[4]{(\gamma-\mu-k_c^2)\mu/9}\;e^{ik_cx}$. Hence, close to the spatial bifurcation, the amplitude of standing waves increases with the fourth root of bifurcation parameter. Experimentally, this bifurcation is abrupt (cf. Fig.~\ref{F-fig2}), hence, as we will discuss below, the dominant nature of noise in this system is multiplicative, that is, the stochastic term is depends on the $A$. Higher order terms considered in model (\ref{E-PDNLS}), like nonlinear dissipation which is neglected in the quasi-reversible limit, can render the spatial bifurcation in an usual supercritical one \cite{PDNLS}, because these terms produce cubic nonlinearities in (\ref{E-DegenerateSupercritical}). Numerically, for negative detuning, the periodic solutions disappear by a saddle node bifurcation close to $\gamma^2=\mu^2 + \nu^2$. Hence, for negative detuning the quiescent state has a subcritical bifurcation that gives rise to pattern state. 

We now turn to the effect of noise on the bifurcation of the parametric surface waves due to the electromagnetically induced vortex flow. To understand how the bifurcation curves presented in Fig.~\ref{F-fig3} are modified we must take into account the fact that Eq.(\ref{E-DegenerateSupercritical}) has been derived through a change of variables close to the identity. Hence, the stochastic terms should be interpreted in the Stratonovich sense
\cite{VanKampen}, which is  equivalently to (Ito sense) 
\begin{eqnarray}
\partial _{t}A &=& (\Gamma+\Delta \tilde{\gamma}) A
-\Delta \tilde{\beta}|A|^2A-(\frac{9}{16\mu}+\Delta
\tilde{\delta})|A|^{4}A
\notag \\
&& +\partial_{xx}A + \sum_{i=0,j=0} \sqrt{\Delta_{i,j}}\zeta_{i,j}
(x,t)A^{i}\bar{A}^{j}, \label{E-Ito}
\end{eqnarray}
where the average over noise realizations satisfies
$\langle\zeta_{i,j}(x,t)A^{i}\bar{A}^{j}\rangle=0$. Hence, the
drift of the system (as it is known the deterministic term in stochastic differential equations~\cite{VanKampen}) has an explicit dependence of the intensity of
noise (terms proportional to $\Delta$). Thus, $\Delta$ and the coefficients
$\{\tilde{\gamma},\tilde{\beta},\tilde{\delta}\}$ related to the modification of Eq.(\ref{E-DegenerateSupercritical}) are functions of the vortex flow, and hence of $I$. $\tilde{\gamma}$ renormalizes the bifurcation threshold,
and $\tilde{\beta}$ characterizes the type of bifurcation: for
positive (negative) $\tilde{\beta}$ the bifurcation is
supercritical (subcritical).  $\tilde{\delta}$ renormalizes the
quintic nonlinearity. In the case that this coefficient is
negative and larger than $9/16\mu$, noise can induce bistability
of parametric surface waves, as observed in Ref.~\cite{FauveResidori2001}. From Fig. \ref{F-fig4} it is clear that
the nonlinear interaction of parametrically amplified waves with 
the vortex flow drastically modifies the bifurcation threshold. 
For instance, for $I$=1.5 A the
bifurcation threshold increases in 20 $\%$ while the saturation exponent is close to 1/2 instead of 1/4. Hence, the
deterministic coefficients, i.e., for $\Delta=0$, are of the same
order of induced stochastic terms. Thus, the bifurcation dynamics is 
driven by the coherent behavior of noise, which gives rise to the effective drift. From the experimental data of Fig.~\ref{F-fig4}, $\Delta\propto I$, which is used to set $\tilde{\gamma}\sim -0.125 A^{-1}$. Notice that $\tilde{\gamma}$ negative as $a_c(I)$ increases with $I$. This allows us to quantify the noise intensity as a function of the vortex flow, as ${\bf F}_L\propto I$. In the case of $\tilde{\beta} \Delta$, although the experimental data does not allow us to compute a precise dependence on $I$, one can observe that $\tilde{\beta} \Delta$ is roughly constant, and thus $\tilde{\beta}$ scales as $I^{-1}$, which sets the cubic term on Eq.~(\ref{E-Ito}) to be of order 1 when the vortex flow is present. Above the critical acceleration $a_c(I)$, the average local amplitude of the parametrically amplified surface waves in presence of the vortex flow is, in good approximation, given by the law $\langle A \rangle_{f_{ex}/2}\sim\epsilon^{1/2}$. Therefore, the induced cubic nonlinearity leads the dynamics close to the bifurcation point, that is, the bifurcation becomes supercritical, in the usual sense.

Quintic supercritical bifurcations are critical instabilities that connect two type of transitions. Inclusions of small stochastic fluctuations disturb the features of the bifurcation. When the intensity of noise is the same order of the deterministic drift, noise changes drastically the threshold, the type, and the bifurcation shape. Hence, noise drives the bifurcation. We have shown that the development of parametrically excited surface waves in the presence of an underlying vortex flow is an example of a noise-driven instability.


The authors thank S. Fauve for fruitful discussions. M.G.C. acknowledges the financial support of ANID-Millenium Science Initiative Program-ICN17$\_$012 (MIRO) and CMM through ANID PIA AFB17000. C. F. acknowledges the financial support of ECOS-ANID C19E05.


\begin{thebibliography}{10}

\bibitem{Lefever} W. Horsthemke and R. Lefever \emph{Noise-Induced Transitions; Theory and Applications in Physics, Chemistry and Biology} (Springer-Verlag, Berlin and New York, 1984) and references therein.

\bibitem{Hangii} L. Gammaitoni, P. H\"anggi, P. Jung, and F. Marchesoni, Rev. Mod. Phys. {\bf 70}, 223 (1998).

\bibitem{OS99}J. Garc\'ia-Ojalvo and J.M. Sancho, \emph{Noise in Spatially
Extended Systems}, (Springer-Verlag, New-York, 1999) and
references therein.

\bibitem{AGTL06} G. Agez, P. Glorieux, M. Taki, and E. Louvergneaux,
Phys. Rev. A \textbf{74}, 043814 (2006).

\bibitem{ClercNoise2006} M.G. Clerc, C. Falcon, and E. Tirapegui, Phys.
Rev. Lett. \textbf{94}, 148302 (2005); Phys. Rev. E {\bf 74}, 011303
(2006).

\bibitem{FDT} H. Nyquist, Phys. Rev. {\bf 32} 110 (1928); 
H.B. Callen and T.A. Welton Phys. Rev. {\bf 83} 34 (1951).

\bibitem{NoiseSupercritical}
G. Agez, M.G. Clerc, and E. Louvergneaux, Phys. Rev. E {\bf 77}, 026218 (2008); I. Ortega, M. G. Clerc, C. Falc\'on and N. Mujica, Phys. Rev. E {\bf81}, 046208 (2010); G. Agez, M. G. Clerc, E. Louvergneaux, and R. G. Rojas, Phys. Rev. E {\bf 87}, 042919 (2013); J. Macías, M. G. Clerc, C. Falc\'on, and M. A. Garc\'ia-\~Nustes, Phys. Rev. E {\bf 88}, 020201(R) (2013).


\bibitem{QuasiReversal} M. Clerc, P. Coullet, and E. Tirapegui,
Phys. Rev. Lett. \textbf{83}, 3820 (1999);
Int. J. Bifurc. Chaos \textbf{11}, (2001) 591;  M.G. Clerc, P. Coullet,
N. Vandenberghe, and E. Tirapegui, Phys. lett. A \textbf{287}, 198 (2001).

\bibitem{FauvePetrelis2005} F. P\'etr\'elis, S. Auma\^itre, and S. Fauve,  Phys.
Rev. Lett. \textbf{94}, 070603 (2005).

\bibitem{PDNLS}
I. V. Barashenkov and E. V. Zemlyanaya, Phys. Rev.
Lett. {\bf 83} 2568 (1999); E. V. Zemlyanaya and N. V. Alexeeva, Theor. Math. Phys. {\bf 159} 870 (2009); I. V. Barashenkov, E. V. Zemlyanaya and T. C. van Heerden, Phys. Rev. E {\bf 83} 056609 (2011).

\bibitem{PDNLSFaraday} W. Zhang and J. Vi\~nals, Phys.
Rev. Lett. \textbf{74}, 690 (1995) and references therein.

\bibitem{Falcon2009} C. Falc\'on and S. Fauve, Physical Review E {\bf80}, 056213 (2009)

\bibitem{Vives} R. Ricou and C. Vives, Int. J. Heat Mass Transfer {\bf 25}, 1579-1588 (1982); A. Cramer, K. Varshney, Th. Gundrum, and G. Gerbeth, Flow Meas. Instrum. {\bf 17}, 1 (2006).

\bibitem{Douady1990} S. Douady, J. Fluid Mech. {\bf 221}, 3383 (1990).

\bibitem{CrossHohenberg} M.C. Cross and P.C. Hohenberg, Rev. Mod.
Phys. \textbf{65}, 851 (1993), and references therein.

\bibitem{PDNLSWater}
J. M. Miles, J. Fluid Mech. 451 {\bf148} (1984); L. Gordillo, T. Sauma, Y. Z\'arate, I. Espinoza, M. G. Clerc, and N. Mujica, Eur. Phys. J. D, {\bf 62}, 39 (2011); L. Gordillo and M. A. Garc\'ia-\~ Nustes, Phys. Rev. Lett. {\bf112}, 164101 (2014); H. Urra {\it et al.},Phys. Rev. E {\bf99}, 033115 (2019); J. F. Mar\'in, {\it et al}, Comm. Phys. {\bf 6} 63 (2023).



Rev. Lett. \textbf{88}, 024502 (2001).


\bibitem{Ursell} T.B. Benjamin, F. Ursell, Proc. R. Soc. London A {\bf 225}, 505 (1954).

\bibitem{Newell} A. Newell, \emph{Solitons in Mathematics and
Physics} (SIAM, Philadelphia, 1985).

\bibitem{NonLinearLatices}  B. Denardo et al., Phys. Rev. Lett. {\bf 68},
1730 (1992).

\bibitem{OpticalFiber}  J. N. Kutz et al., Opt. Lett. {\bf 18}, 802 (1993).

\bibitem{OPO}  S. Longhi, Phys. Rev. E {\bf 53}, 5520 (1996).

\bibitem{Magnetization}  I. V. Barashenkov, M. M. Bogdan, and V. I. Korobov,
Europhys. Lett. {\bf 15}, 113 (1991).

\bibitem{ClercCoulibalyLaroze2008} M.G. Clerc, S. Coulibaly, and D. Laroze,
Phys. Rev. E {\bf 77}, 056209 (2008).


\bibitem{ChainPendula}  N.V. Alexeeva, I.V. Barashenkov, and G.P. Tsironis,
Phys. Rev. Lett. {\bf 84}, 3053 (2000).

\bibitem{CoulletFrischSonnino} P. Coullet, T. Frisch, and G. Sonnino,
Phys. Rev. E {\bf 49}, 2087 (1994).

\bibitem{ClercFalconTirapegui} M.G. Clerc, C. Falcon, and E. Tirapegui,
Phys. Rev. E {\bf 74}, 011303 (2006).

\bibitem{VanKampen}  N.G. van Kampen, {\em Stochastic processes in physics
and chemistry} (Elsevier North-Holland, Amsterdam, 1981).


\end{thebibliography}
\end{document}